\documentclass{article}
\usepackage{arxiv}
\usepackage[utf8]{inputenc} % allow utf-8 input
\usepackage[T1]{fontenc}    % use 8-bit T1 fonts
\usepackage{hyperref}       % hyperlinks
\usepackage{url}            % simple URL typesetting
\usepackage{booktabs}       % professional-quality tables
\usepackage{amsfonts}       % blackboard math symbols
\usepackage{nicefrac}       % compact symbols for 1/2, etc.
\usepackage{microtype}      % microtypography
\usepackage{lipsum}		% Can be removed after putting your text content
\usepackage{doi}
\usepackage{array}
\usepackage{enumitem}
\usepackage{graphicx}
\usepackage[square,sort,comma,numbers]{natbib}

\title{Monitoring and Evaluating Astronomy-for-Development Initiatives}

\author{ \href{https://orcid.org/0000-0002-9745-0504}{\includegraphics[scale=0.06]{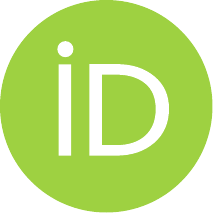}\hspace{1mm}Joyful E. Mdhluli}{ on behalf of the IAU Office of Astronomy for Development} \thanks{Visit our website, www.astro4dev.org or email: info@astro4dev.org} \\
	International Astronomical Union's Office of Astronomy for Development\\
	South African Astronomical Observatory\\
        Cape Town, South Africa\\
	\texttt{joy@astro4dev.org} \\
	\And
}

\renewcommand{\shorttitle}{\it{Monitoring and Evaluating Astro4Dev intiatives}}

\hypersetup{
pdftitle={Monitoring and Evaluating Astronomy-for-Development Initiatives},
pdfsubject={astro-ph.IM},
pdfauthor={Joyful E.~Mdhluli},
pdfkeywords={Astronomy, Sustainable development, Impact, Monitoring, Evlauation},
}

\begin{document}
\maketitle

\begin{abstract}
This paper serves as a practical guide for individuals and organisations seeking to design, implement, and evaluate astronomy-for-development initiatives, as well as those preparing proposals for the International Astronomical Union’s Office of Astronomy for Development (IAU OAD) annual Call for Proposals. The paper aims to  outline how systematic evidence collection can strengthen project design, enhance accountability, and increase the likelihood of measurable impact. It explains the distinction between monitoring and evaluation, provides guidance on when and how evaluation should be undertaken, and summarises key evaluation types - process, feasibility, impact, outcome, economic, and summative, relevant to astronomy-based interventions. In addition to conceptual discussion, the paper presents a set of practical steps, reflective questions, and examples to help project teams develop a clear theory of change, define appropriate indicators, and anticipate risks and unintended consequences. By contextualising M\&E within the broader goals of the OAD and the Sustainable Development Goals (SDGs), this work aims to empower practitioners to create evidence-informed, community-driven, and sustainable astronomy-for-development projects that deliver both local and global benefit.
\end{abstract}

% keywords can be removed

\section{Introduction}
The global movement to use astronomy as a tool for sustainable development has gained momentum over the past decade, driven by the work of the International Astronomical Union’s Office of Astronomy for Development (IAU OAD). By connecting the inspiration of the cosmos with the practical needs of communities, astronomy-for-development projects have demonstrated potential to advance education, equality, innovation and socio-economic wellbeing \cite{comment1}-\cite{comment4}. Yet, as the field matures, a central challenge has emerged: how can we measure and demonstrate the real-world impact of these interventions? Without clear evidence, even the most inspiring initiatives risk remaining anecdotal, their outcomes difficult to replicate, scale, or justify to funders and policymakers.

Monitoring and Evaluation (M\&E) provide the mechanism through which such impact can be systematically understood. Within development practice more broadly, M\&E ensures that projects are not only well-implemented but also relevant, effective, efficient and sustainable. Applied to astronomy-for-development, M\&E plays an equally vital role in refining project design, ensuring accountability, identifying unintended consequences, and fostering organisational learning. However, astronomy projects often face distinctive obstacles in this regard: they operate across disciplines, involve intangible outcomes such as inspiration or curiosity, and are frequently implemented by small teams with limited evaluation capacity.

In response to these challenges, the IAU OAD has developed an M\&E framework that adapts principles from the broader development sector to the specific realities of astronomy-based interventions. The framework distinguishes between monitoring, the continuous tracking of activities, progress and emerging issues, and evaluation, which systematically investigates a project’s effectiveness, feasibility, and longer-term impact. It also recognises that evaluation is not always appropriate or feasible, especially for small or exploratory projects, and that the choice of evaluation type such as process, feasibility, outcome, impact, economic or summative, should depend on project scope, purpose and available resources.

This paper situates the OAD’s approach to monitoring and evaluation within the wider context of evidence-based development practice. It explores how a clear theory of change, appropriate indicators, and participatory evaluation can strengthen astronomy’s contribution to the Sustainable Development Goals (SDGs). By analysing the conceptual foundations and practical guidance of the OAD’s M\&E model, this work seeks to contribute to a growing discourse on accountability, learning, and impact measurement in science-for-development initiatives. Ultimately, it argues that robust M\&E is essential not only to justify investment in astronomy-for-development, but to ensure that its outcomes are meaningful, equitable, and sustainable for the communities it aims to serve.

\subsection{Purpose and Structure}
The primary aim of this paper is to support practitioners, educators, and community organisers who wish to design and implement astronomy-for-development initiatives, as well as those seeking to submit proposals to the IAU Office of Astronomy for Development (OAD) \cite{cfp}. To achieve this, the paper is structured around three complementary objectives. First, it introduces the conceptual foundations of Monitoring and Evaluation (M\&E) in the context of astronomy-for-development, clarifying key distinctions between monitoring and evaluation and highlighting when each is appropriate. Second, it outlines practical guidance for developing a theory of change, selecting measurable indicators, and choosing appropriate evaluation types, drawing on the OAD framework and illustrative examples. Third, it presents a set of reflective questions and actionable steps aimed at helping project teams anticipate challenges, assess feasibility, and design interventions that are community-driven, evidence-informed, and sustainable. This structure ensures that readers can not only understand the principles behind effective M\&E but also apply them directly to the planning, execution, and proposal development of astronomy-for-development projects.

\section{Basic Understanding of Monitoring and Evaluation}
Research has shown that interventions often run into unexpected barriers and can produce unexpected negative, as well as positive, impacts \cite{chapman}. As a result, non-profit organisations and other intervention providers have come under increasing pressure from funders and other stakeholders to provide information about their performance. Monitoring and evaluation refer to a combination of activities and procedures used to effectively measure, report and learn from project performance.

Monitoring and evaluation are considered good practice for organisations engaged in social intervention because they enable providers to:
\begin{itemize}
    \item Determine whether a project successfully produced desired outcomes.
    \item Understand how a project worked or did not work.
    \item Check for unforeseen negative consequences that require attention in future.
    \item Identify which activities produce the largest or most important positive impacts.
    \item Identify which strategies are most cost-effective (i.e. resource efficient) in achieving project goals.
    \item Assess how meaningful, sustainable , accessible and relevant a project was for participants and other stakeholders.
    \item Learn from experience how to improve and build on project experience.
    \item Demonstrate impact to others.
\end{itemize}

\textbf{Monitoring} refers to indicators used throughout the life of a project to measure progress and record how a project was designed, implemented and delivered. Monitoring data capture whether a project is implemented in a way that is consistent with its design and whether any unanticipated barriers were encountered (for example, no one applied to take part in the project). The monitoring phase of a project also provides a framework through which to track progress and identify problems early on, thus offering a way through which corrective action and improvements can be made ‘on the ground’ and recorded without compromising project objectives.

\textbf{Evaluation} focuses on “what works”, as well as how, for whom and under what conditions. The key difference between evaluation and monitoring is that evaluation is about using data, including monitoring and post-project outcome data, in order to draw conclusions about project impact. Different types of evaluation are used to address different dimensions of project impact and effectiveness (see the next section for a summary of evaluation types). Different research designs are used for different types of evaluation; and different types of evaluations are of interest to different stakeholders. It is considered good practice to conduct an impact or outcome evaluation in combination with an economic evaluation wherever possible; however, these two types of evaluation are also the most technically challenging.

\section{Types of Evaluation}
Evaluation can take several forms, each serving a distinct purpose. Listed below are different evaluation types that can be implemented in astronomy-for-development initiatives. 
\begin{table}[htbp]
\centering
\begin{tabular}{l >{\raggedright\arraybackslash}p{10cm}}
\toprule
\textbf{Evaluation Type} & \textbf{Focus Questions} \\
\midrule
Mechanistic/Theoretical & 
\begin{minipage}[t]{10cm}
\begin{itemize}[leftmargin=*]
\item What is the project’s theory of change? i.e. how and why does a project lead to changes in desired (and/or secondary) outcomes?
\item How does the project’s design and implementation reflect that theory?
\item Which intermediate outcomes are produced during the project?
\item Which project components cause intermediate or final changes in outcome?
\item Which individuals (or communities) are likely to respond best and why?
\item Are there any possible negative outcomes? why do these occur? How can these risks be minimised or avoided?
\end{itemize}
\end{minipage} \\
\midrule
Process & 
\begin{minipage}[t]{10cm}
\begin{itemize}[leftmargin=*]
\item How was the project implemented?
\item Which aspects were feasible and which aspects were problematic?
\item How did project stakeholders experience it?
\item Which components were effective? Which were not?
\item How did the project produce outcomes and for whom? I.e. were theories of change shown to be correct or do these theories require adjustment?
\end{itemize}
\end{minipage} \\
\midrule
Feasibility & 
\begin{minipage}[t]{10cm}
\begin{itemize}[leftmargin=*]
\item Would it be possible to scale the project up and deliver on a much larger scale?
\item Is it possible to replicate the project?
\item Has the project gained enough stakeholder support to sustain or expand it?
\item Are impacts long-lasting?
\end{itemize}
\end{minipage} \\
\midrule
Impact/Outcome & 
\begin{minipage}[t]{10cm}
\begin{itemize}[leftmargin=*]
\item Did the project meet the overall needs?
\item Was any change significant and was it attributable to the project?
\item How valuable are the outcomes to the organisation, other stakeholders, and participants?
\end{itemize}
\end{minipage} \\
\midrule
Economic & 
\begin{minipage}[t]{10cm}
\begin{itemize}[leftmargin=*]
\item Was the project cost effective? i.e. do outcome gains equal or exceed investments?
\item Was there another alternative that may have produced the same or better outcomes while using fewer resources?
\end{itemize}
\end{minipage} \\
\midrule
Summative & 
\begin{minipage}[t]{10cm}
\begin{itemize}[leftmargin=*]
\item What lessons have been learned? How can future projects build on this project? Which questions require further research?
\end{itemize}
\end{minipage} \\
\bottomrule
\end{tabular}
\caption{Evaluation Types and Focus Questions}
\end{table}
 
\section{Defining Key Terms}
Astronomy for development (Astro4Dev) is concerned with activities that involve people rather than stars. That is to say, Astro4Dev projects are projects that seek to affect human development, not achieve scientific objectives. Astro4Dev projects are thus classed as “social interventions”: interventions, policies, practices or programmes that seek to improve social welfare outcomes by addressing social, economic, health, psychology, education or behaviour problems.

Some diverse examples of social interventions include humanitarian relief after natural disasters; HIV prevention campaigns; ‘sin’ taxes on cigarettes; reductions in classroom sizes to improve education outcomes; drug treatment in prisons to reduce reoffending; and after-school clubs to improve childhood school performance and reduce delinquency. Even though the ‘social problems’ that they address are not always made explicit, all of the OAD projects conducted to date can be classes as social interventions too.

Evaluation is talked about a lot across fields of social intervention. Everyone seems to mean something slightly different. There is no consensus on a technical definition. The Gates Foundation uses the following definition \cite{GatesFoundation}:

\textit{Evaluation is the systematic, objective assessment of an ongoing or completed intervention, project, policy, program, or partnership. Evaluation is best used to answer questions about what actions work best to achieve outcomes, how and why they are or are not achieved, what the unintended consequences have been, and what needs to be adjusted to improve execution.}

There are two main types of evaluation, focused on different questions listed in the definition above:
\begin{itemize}
    \item Impact (aka Summative) Evaluation, which focuses on whether and to what extent target outcomes were achieved (i.e. measuring impact)
    \item Process (aka Formative) Evaluation, which focuses on how, for whom and under what conditions a project works

\end{itemize}
Evaluation is typically paired with monitoring (hence the phrase “Monitoring and Evaluation”, sometimes abbreviated to “M\&E”).

Monitoring focuses on determining how a project was delivered and whether this deviated from its original plan or design. Monitoring can be used to identify obstacles to implementation (e.g. under-budgeting; the need for more staff training etc.) and ensure accountability (e.g. reduce misallocation of funds or even deliberate corruption in aid organisations).

\begin{figure}[h!]
    \centering
    \includegraphics[width=\textwidth]{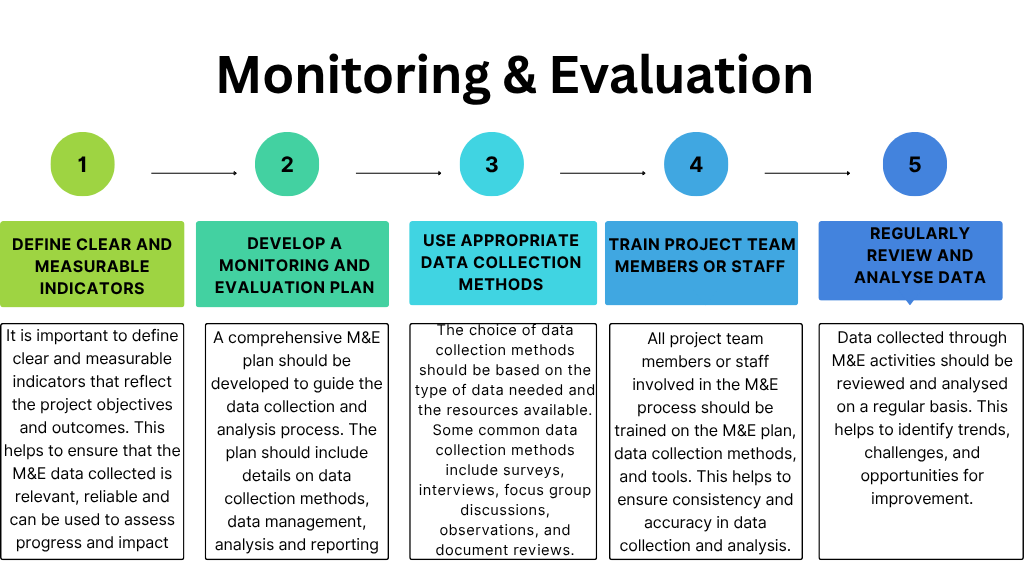}
    \caption{Practical steps to Monitoring and Evaluation.}
    \label{monitoring}
\end{figure}

The results of both kinds of evaluation (impact and process) and of project monitoring are combined to provide lessons for future projects and improve practice over time. Beyond understanding these conceptual differences, effective M\&E also depends on applying practical steps and best practices that strengthen the reliability, usefulness, and impact of the process. For example, in development initiatives, the following practices can help ensure meaningful monitoring and evaluation, see Fig. \ref{monitoring}:
\begin{itemize}
    \item Developing M\&E framework: The framework must outline the key indicators that will be used to measure the progress towards project goals or outcomes, as well as the data collection methods and tools that will be used to gather this information. 
    \item Collecting data: Data can be collected via a variety of methods, these include surveys, interviews and observations. The data collected could be both qualitative and quantitative so as to provide a comprehensive understanding of the projects’ impact and outcomes. 
    \item Analysing data: Once the data has been collected, it needs to  be analysed and interpreted, to identify if the project has met its objectives or outcomes. If there are patterns and trends in the data that can be drawn out to support the reporting. This information can be used to assess the effectiveness of project activities and to identify areas where improvements can be made. 
    \item Reporting findings: The results of the M\&E process can be communicated to project stakeholders who are involved, these may include the funders, project team members and staff if there is. The reporting sort of builds support for the project and ensures that it remains accountable to its goals, objectives and set outcomes.
\end{itemize}

\section{Reasons to Evaluate}
Evaluations are often required by funders (e.g. by nearly all government funds for education and development projects). There are, however, many more reasons to conduct evaluations. These include:
\begin{itemize}
    \item \textbf{Cost-effectiveness} Determining whether a project impacted intended outcomes and the size of those impacts can be used to select between different projects, ensuring that limited resources are allocated to those projects that are most effective
    \item \textbf{Unintended consequences} Human beings are complex and social systems are even more so. Social initiatives conducted over the past century (and most people’s personal experience!) provide substantial evidence of how even the simplest interventions and efforts to help others can have significant and far-reaching unintended consequences. Sometimes these are good but it is surprisingly easy to cause inadvertent harm. Evaluation provides a way to check for harm and mitigate the risk of repeating and/or ‘scaling up’ an approach that causes unforeseen harm.
    \item \textbf{Improve practice over time} Another consequence of social complexity is that it can be difficult to design a project that has a large impact. It usually takes time and a ‘learning cycle’ to arrive at an approach that really works. Evaluation enables lessons about what works to be used and incorporated into future practice so that project designs improve and become more effective over time (this is the learning cycle embedded in the \href{https://astro4dev.org/funded-projects/impact-cycle/}{OAD Impact Cycle}).
    \item \textbf{Contribute to knowledge and practice} Evaluation offers lessons for others about what went well, what worked and what did not. Since evaluation findings can be shared, evaluations can be used to improve not only OAD project designs and practice but also a broader understanding of what works in related fields (e.g. effective techniques for science communication); and to improve the design of similar projects conducted by other organisations and actors (e.g. other science unions, AstroEdu, UNAWE etc.).
    \item \textbf{Demonstrating impact to stakeholders} For a project to be sustainable, it must be supported by all key stakeholders. These include the people who deliver the project (e.g. OAD project leaders and teams); those who fund the project (e.g. the OAD, the IAU, Kickstarter donations etc.); those who participate (e.g. students, teachers); and any other affected communities or parties (e.g. school districts). Evaluations provide project supporters with evidence that they take their project objectives seriously and prioritise achieving positive outcomes (i.e. by taking a risk with an evaluation that could show that the project does not work). If an evaluation shows positive results (or once a project has been improved to the point where it shows positive results), the project supporters can use these results to build trust with target participants and attract support for its continuation.
    \item \textbf{Increasing funding and scale of delivery} Most large international funders will not allocate significant resources to any intervention that has not been demonstrated to bring about positive outcomes. Organisations that require rigorous impact evaluation before scaling-up include The Gates Foundation, the largest private philanthropic foundation in the world; UN agencies; Oxfam; and most government aid departments such as USAID and the UK’s DFID).
\end{itemize}

Evaluations are particularly important to conduct when projects are
\begin{itemize}
    \item Using innovative methods that have not been tested.
    \item Operating in a domain where knowledge is lacking about what works.
    \item Working on problems whose mechanisms are poorly understood.
    \item Attempting to change behaviours, attitudes or social structures.
    \item Trying to affect outcomes that are difficult to observe.
\end{itemize}
For example, we hope that an OAD project proposal may focus on improving gender inclusion in the sciences. This project would probably be a good candidate for having an in-built evaluation. This is because we know that there are multiple interacting causes for disparities in gender representation in the sciences and that these causes are likely to vary over time and across contexts. We also don’t know of any highly effective solutions to these causes; and probably do not know what all the causes are. Furthermore, we do know that there have been several high-profile examples of projects that sought to enhance girls’ entry into science and failed to do so or actually had the opposite effect (e.g. ‘Science It’s a Girls Thing’). An evidence-based design (incorporating what is known) combined with a carefully designed evaluation framework would help mitigate the risk of causing unforeseen harm and ensure that the project contributed to evolving knowledge and practice in this area - at best, the project would be found to be effective and could be used by others. At worst, the project would contribute to our understanding of what does (and does not) work and thus help future projects design more effective approaches.

\section{Constraints on Evaluation}
Evaluation is not always appropriate and ideal evaluations are rarely possible. Evaluations are likely to be particularly limited for smaller projects, where a huge expensive measurement exercise would end dwarfing the costs of the project.

In determining whether and how to evaluate a project, the following constraints should be kept in mind:
\begin{itemize}
    \item \textbf{Cost:} International development agencies, funders and organisations typically allocate approximately 10-20\% of each project’s budget to monitoring and evaluation. While this might seem like a lot, evaluation can enable much larger gains over time by ensuring that the most effective projects are identified and scaled up while the most ineffective (and any harmful!) projects are abandoned.
    \item \textbf{Burden on implementation:} This is a particularly serious concern for OAD projects, which are often delivered by volunteers in their spare time. In the ideal case, evaluation design and data collection would be conducted by a separate (independent) team. This has the advantage of reducing the burden on those implementing the project; and reducing the risk of conflicting interests (since those who designed or implement a project are likely to be invested in its success!). In reality, it is often impossible to budget for or engage independent evaluators.
    \item \textbf{Restricting outcomes:} When outcomes are difficult to measure, there is a risk that attention will be paid only to what is measurable. If measures are poorly designed or inadequate, this can create problems; even when measures are good, they must necessarily focus on a narrow selection of outcomes. There may be others of interest that are then ignored. For example, the selection of science \& research indicators may influence policymakers to focus on specific areas of science output that are easily (or currently) measured rather than on those that could be done; or that have more restricted (local) relevance (see a blog on this topic by the LSE).
    \item \textbf{Political/personal/emotional:} Constraints on evaluation means that (if an evaluation is appropriate) monitoring and evaluation activities should be embedded in the project and data collection burdens minimized. This can mean streamlining reporting and auditing requirements; limiting the number of outcomes measured; and limiting the kind of data collected (e.g. level of detail about participants). Technological solutions should also be considered wherever possible.
\end{itemize}
It is also important to be aware that these constraints also mean that a single evaluation is unlikely to answer all questions about a project. For example, the first randomised controlled trial (RCT) conducted by the OAD was an impact evaluation of in-school astronomy activities for children. To limit costs and implementation burdens, only 5 questions were asked: all of these questions concerned the children’s social identities and attitudes to in-group and out-group members. These questions were based on the project’s core theory. While the results did not support the theory, there were many unanswered questions: did the activities increase interest in science? did the students’ intergroup attitudes change in other ways, not captured by those questions? It could be that the project had positive effects in domains that were simply not measured! The experiment will thus need to be repeated with different outcome measures that capture all of the ‘intended outcomes’ that the OAD community expects - and with more open-ended explorations to identify possible unexpected outcomes! - before we can reach conclusions about whether it ‘worked’ in a broader sense.

\section{When not to Evaluate}
Not all Astro4Dev projects need to or even should be evaluated, for example projects that:
\begin{itemize}
    \item Are not clearly defined (replicable project).
    \item Are too complex: evaluate replicable components, not complex combinations.
    \item Have no aims to change observable outcomes (e.g. projects aimed to improve inspiration, which is not an observable outcome).
    \item Test no clear hypothesis.
\end{itemize}
Evaluation is a low priority when the results of our efforts are easily observable. It is also a low priority when our projects are conducting basic scientific research, developing but not distributing products or tools, or creating new data sets or analyses. In such cases, our projects’ self-reported progress data and existing protocols (such as for clinical trials) provide sufficient feedback for decision making and improvement.

\section{Conclusion}
This paper reinforces that monitoring and evaluation are not optional add-ons but integral components of effective astronomy-for-development practice. By differentiating between monitoring (tracking implementation fidelity and emerging barriers) and evaluation (assessing what works, for whom, under what conditions), the IAU OAD guidance frames M\&E as a mechanism for both accountability and learning. In the context of astronomy-based interventions, the challenges are many: resource constraints can limit rigorous outcome measurement; intangible goals (such as inspiration or attitude change) are hard to operationalise; and projects may inadvertently produce negative or inequitable effects if not carefully monitored. The guidance’s caution about “when not to evaluate” is particularly valuable: small pilot projects or those with purely inspirational aims may not justify costly evaluations. Yet for interventions seeking behavioural change, skills transfer, or socio-economic outcomes, embedding evaluation early - defining a clear theory of change, selecting appropriate indicators, planning for cost-effectiveness - enables better decision-making, scale-up potential and transferable learning across contexts. In summary, astronomy can contribute to human development, but only if projects are designed with strong M\&E frameworks that ensure meaningful, relevant, accessible and sustainable outcomes. As the domain matures, practitioners should move beyond output counting, embrace mixed-methods evaluations, and share evidence of what works (and what doesn’t). Future work should prioritise longitudinal studies, economic evaluations and comparative research across projects and regions. In doing so, we will strengthen the credibility of astronomy-for-development and its potential to align with and support the SDGs.

\end{document}